\documentclass[a4paper,11pt]{article}
\usepackage{pos}
\usepackage{xspace}

\newcommand{\mtpole}{\ensuremath{m_{t}^{\text{pole}}}\xspace}
\newcommand{\ttbar}{\ensuremath{t\overline{t}}\xspace}
\newcommand{\mtt}{\ensuremath{M(\ttbar)}\xspace}
\newcommand{\ytt}{\ensuremath{|y(\ttbar)|}\xspace}

\title{
Compatibility between theoretical predictions and experimental data for top-antitop hadroproduction at NNLO QCD accuracy
}
\ShortTitle{Compatibility theory predictions/experimental data for \ttbar $+ X$ hadroproduction at NNLO QCD accuracy}

\author*[a]{Maria Vittoria Garzelli}
\author[b]{Javier Mazzitelli}
\author[a]{Sven-Olaf Moch}
\author[a]{Oleksandr Zenaiev}

\affiliation[a]{Universit\"at Hamburg, {II}. Institute for Theoretical Physics,\\
Luruper Chaussee 149, 22761 Hamburg, Germany}
\affiliation[b]{
Paul Scherrer Insstitut, CH–5232 Villigen, Switzerland}

\emailAdd{maria.vittoria.garzelli@desy.de}
\emailAdd{javier.mazzitelli@psi.ch}
\emailAdd{sven-olaf.moch@desy.de}
\emailAdd{oleksandr.zenaiev@desy.de}

\abstract{ 
We compare double-differential normalized production cross sections for top-antitop $+ X$ hadroproduction at NNLO QCD accuracy, as obtained through a customized version of the \texttt{MATRIX} framework interfaced to \texttt{PineAPPL}, with recent data by the ATLAS and CMS collaborations.

We take into account theory uncertainties due to scale variation and we see how predictions vary as a function of parton distribution function (PDF) choice and top-quark pole mass value, considering different state-of-the-art PDF fits with their uncertainties.

Notwithstanding the overall reasonable good agreement, we observe discrepancies at the level of a few $\sigma$'s between data and theoretical predictions in some kinematical regions, which can be alleviated by refitting the top-quark mass value, and/or the PDFs and/or $\alpha_s(M_Z)$, considering the correlations between these three quantities.

In a fit of top-quark mass standalone, we notice that, for all considered PDF + $\alpha_s(M_Z)$ sets used as input, some datasets point towards top-quark pole mass values lower by about $2 \sigma$'s than those emerging from fitting other datasets, suggesting a possible tension between experimental measurements using different decay channels, and/or the need of better estimating uncertainties on the latter.
}

\FullConference{The European Physical Society Conference on High Energy Physics (EPS-HEP2023)\\
 21-25 August 2023\\
Hamburg, Germany\\}

\begin{document}
\maketitle

\section{Introduction}
\label{intro}

Top-quark measurements at the Large Hadron Collider (LHC) play a crucial role in modern particle physics in order to precisely extract key parameters of the Standard Model (SM). Furthermore, they provide insights into the electroweak symmetry breaking mechanism and are a vital component of searches for physics beyond the SM.

The present study aims at the determination of the top-quark mass, here extracted from a comparison of inclusive and differential cross-section data for $t\bar{t}+X$ production collected by the LHC experiments ATLAS and CMS with theoretical predictions including higher-order corrections in quantum chromodynamics (QCD) and computed for the top-quark mass in the on-shell renormalization scheme, \mtpole. We use a customized version of \texttt{MATRIX}~\cite{Catani:2019hip, Grazzini:2017mhc}, optimized for the $pp \rightarrow t\bar{t} + X$ process and interfaced to \texttt{PineAPPL}~\cite{Carrazza:2020gss}, for the computation of all NNLO QCD theory predictions with NNLO QCD uncertainties (without utilizing $K$-factors and any further approximation linking NNLO results to lower-order ones). These are the first fits of \mtpole with exact NNLO accuracy using LHC double-differential $t\bar{t}~+~X$ data, to the best of our knowledge.

\section{Modified \texttt{MATRIX}+\texttt{PineAPPL} framework for theoretical calculations}

We started from the \texttt{MATRIX} version used for the computations presented in Ref.~\cite{Catani:2019hip} and we performed a number of optimizations in the program flow and execution. A few per mill accuracy requires the generation of various billions of $t\bar{t} + X$ NNLO events, which takes $\mathcal{O}(10^5)$ CPU hours. A general solution to this problem is to use interpolation grids, where partonic matrix elements are stored in such a way that they can be convoluted later with any (PDF + $\alpha_s(M_Z)$) set. We choose the \texttt{PineAPPL} library~\cite{Carrazza:2020gss} which is capable of generating grids and dealing with them in an accurate way to any fixed order in the strong and electroweak couplings, and which supports variations of the renormalization and factorization scales, $\mu_r$ and $\mu_f$. 

In order to validate our implementation of the interface to \texttt{PineAPPL}, we compared the genuine theoretical predictions from \texttt{MATRIX} with those obtained using the \texttt{PineAPPL} interpolation grids and found them to agree within a few per mill. We also compared our theoretical predictions with those from Ref.~\cite{Czakon:2016dgf} and found them to agree within the $\approx 1 \%$ uncertainties of the latter. Based on these validation studies, we assign a $1\%$ uncorrelated uncertainty in each bin of our predictions. 

\section{NNLO fits of the top-quark pole mass value} 
\label{sec:fit}

\begin{figure}[htb]
    \centering
    \includegraphics[width=0.78\textwidth,clip,trim=0 41 0 6.5]{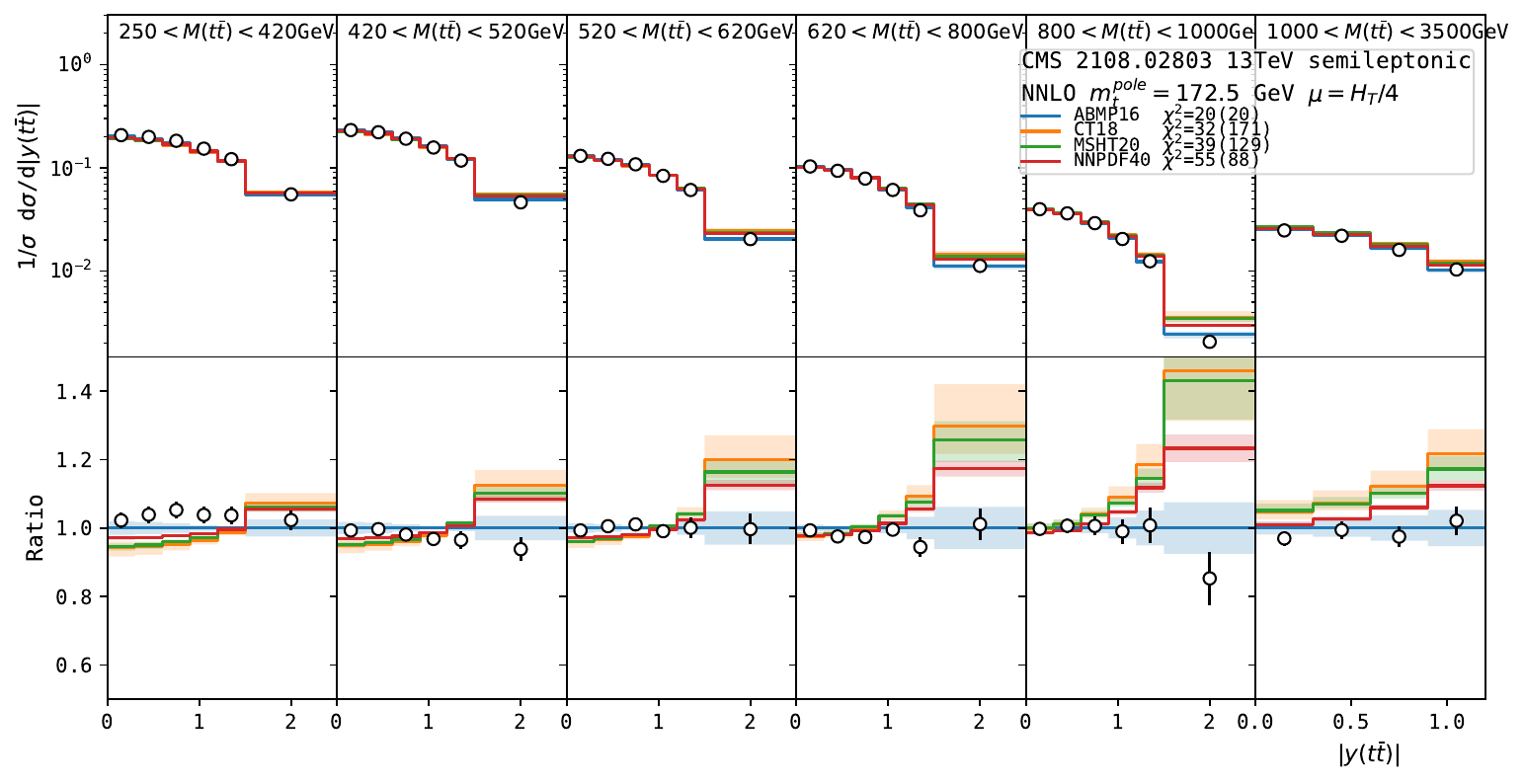}
    \includegraphics[width=0.78\textwidth,clip,trim=0 41 0 6.5]{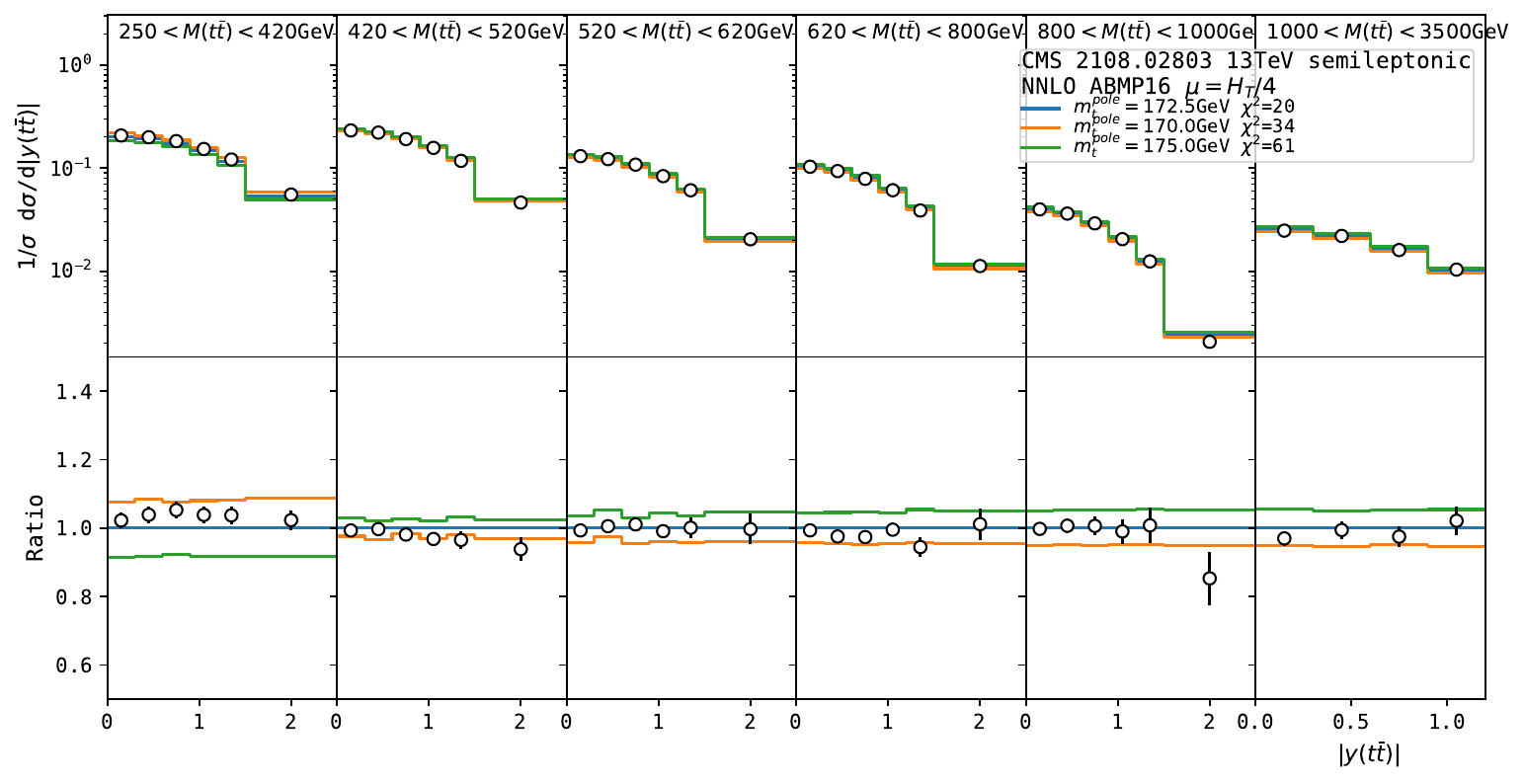}
    \includegraphics[width=0.78\textwidth,clip,trim=0 8 0 6.5]{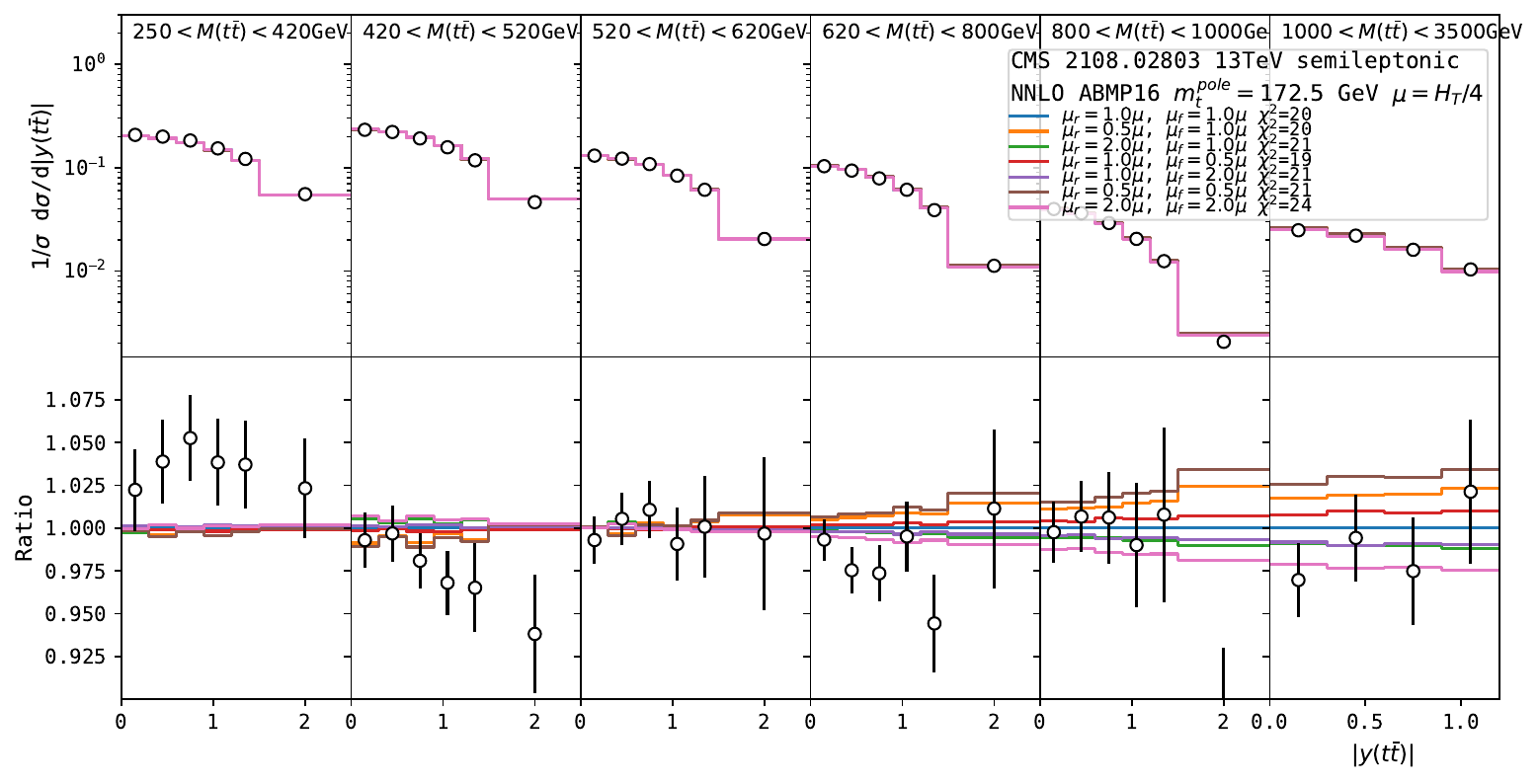}
    \caption{Comparison of the experimental data
from Ref.~\cite{top20001}
to the NNLO predictions 
obtained using different PDF sets (upper), and, for the ABMP16 central PDF member, different \mtpole values (middle) and different $\mu_R$, $\mu_F$ scales varied by factor of two around the central value $\mu_{R,0} = \mu_{F,0} = \mu_0 = H_T/4$ (lower).}
    \label{fig:top20001}
\end{figure}

For our analysis, we use measurements of the absolute total and normalized differential inclusive \ttbar~+~$X$ cross sections. We collect all available and up-to-date ATLAS and CMS measurements of total \ttbar + $X$ cross sections which appear on the summary plot of the total \ttbar + $X$ cross sections by the LHC Top working group as of June 2023~\cite{lhctopwg}. For differential measurements, we choose cross sections as a function of the invariant mass of the \ttbar pair, \mtt, and, if available, double-differential cross sections as a function of \mtt and rapidity \ytt of the \ttbar pair~\cite{top20001,top18004,a190807305,a200609274,top14013,a151104716,a160707281,a14070371}.
We use four state-of-the-art NNLO proton PDF + $\alpha_s(M_Z)$ sets as input of the theory computations: ABMP16~\cite{Alekhin:2017kpj}, CT18~\cite{Hou:2019efy}, MSHT20~\cite{Bailey:2020ooq} and NNPDF4.0~\cite{NNPDF:2021njg}.

\begin{figure}[htb]
    \centering
    \includegraphics[width=0.475\textwidth]{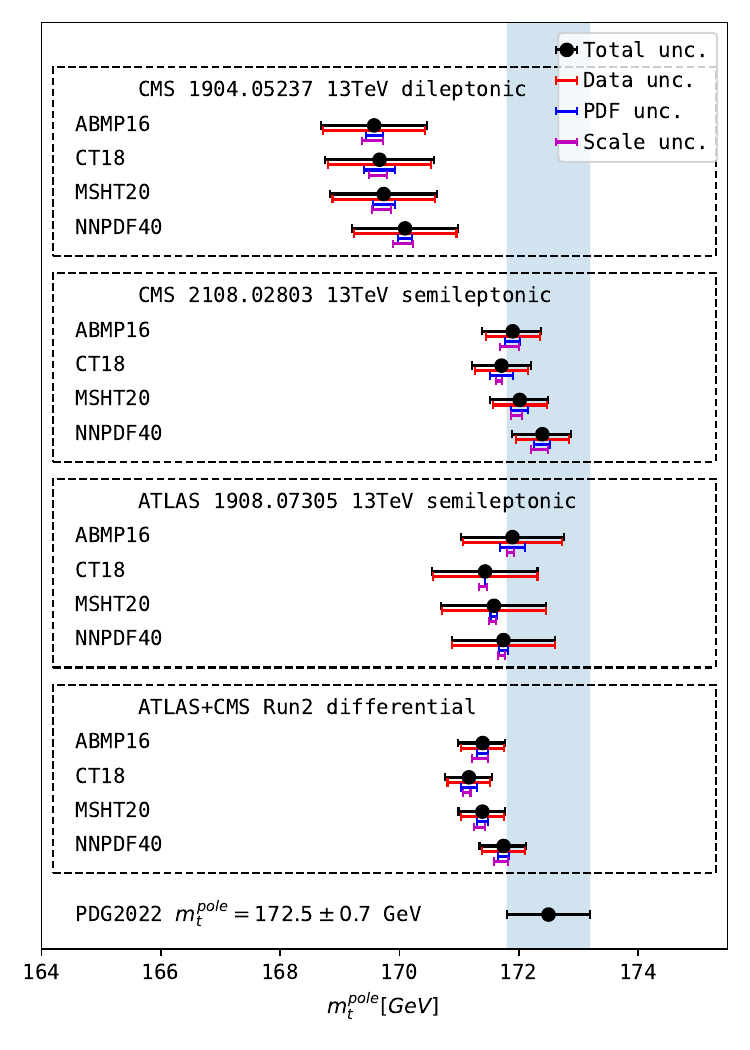}
    \includegraphics[width=0.475\textwidth]{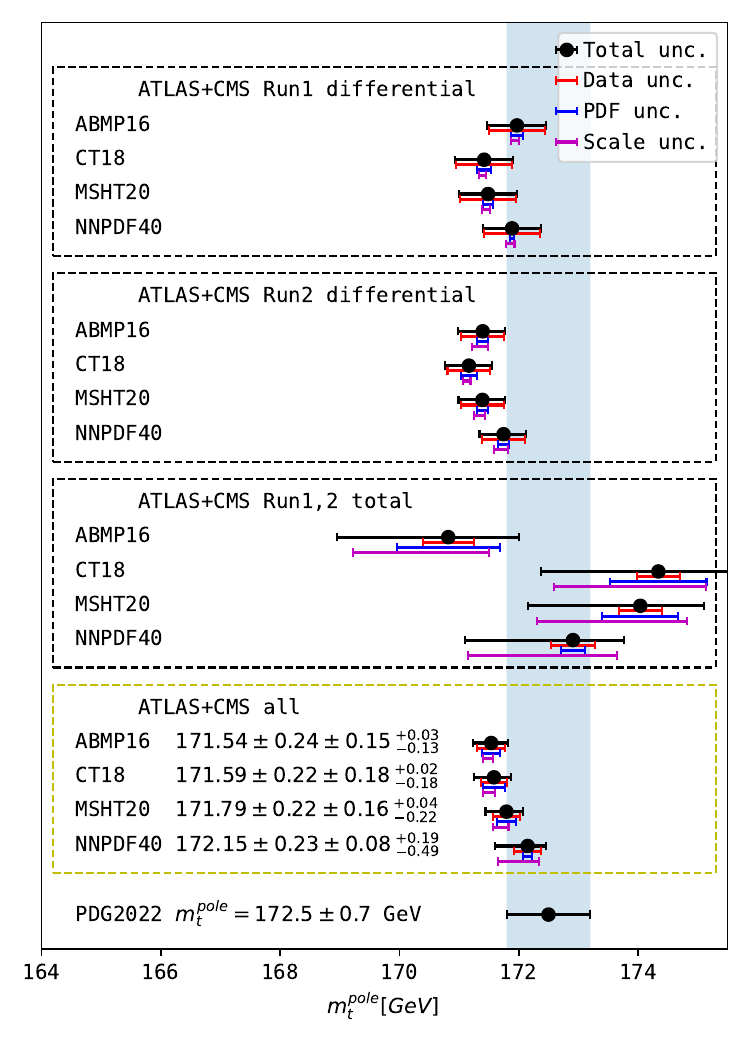}
    \caption{Summary of the \mtpole values extracted from Run2 measurements of differential cross sections (left), and Run1 and Run2 measurements of differential and total \ttbar $+ X$ cross sections (right). We assign as scale uncertainty the maximum difference~on \mtpole from the 7-point ($\mu_R$, $\mu_F$) variation.     }
    \label{fig:mt}
\end{figure}

As an example of data-to-theory comparison, in Fig.~\ref{fig:top20001} the rapidity distribution of the $t\bar{t}$-quark pair, \ytt, is plotted in various $t\bar{t}$ invariant mass \mtt bins, corresponding to different panels, and compared to the experimental data of Ref.~\cite{top20001}, a state-of-the-art CMS analysis with $t\bar{t}$-quark pairs decaying in the semileptonic channel.
From the first row of plots, it emerges that the best agreement between theoretical predictions and experimental data as for the shape of the distributions, that is probed when considering normalized cross-sections, is achieved when using the ABMP16 PDFs. Predictions with the CT18, MSHT20 and NNPDF4.0 PDFs show a similar trend among each other, but shapes systematically different from those of the experimental distributions at large \ytt, overestimating them. 
The plots of the second row, all obtained with the ABMP16 PDFs, show that the largest is \mtpole, varied in the range 170~GeV $< m_t^{\mathrm{pole}} < 175$~GeV, the smallest is the cross-section for low \mtt close to threshold, while the opposite is true for large $M_{t\bar{t}}>420$~GeV because of the cross-section normalization. 
The plots of the third row show the behaviour of the \ytt distribution under ($\mu_R$, $\mu_F$) variation. Scale uncertainties increase at large \mtt, reaching values up to ~$\pm$~3\%  in the largest \mtt bin, an amount comparable to data uncertainties in this kinematic region.

The \mtpole fits were then performed using the \texttt{xFitter} framework~\cite{Alekhin:2014irh}, an open source QCD fit framework.
In Fig.~\ref{fig:mt} the results for the \mtpole extraction from various ex\-pe\-ri\-men\-tal datasets are shown. In the left panel, results related to individual Run~2 measurements of differential \ttbar~+~$X$ cross sections are shown separately, while the right panel reports the results of the global fit, including both total and differential cross-section Run 1 and 2 data.
The results of the extraction using either differential or total cross sections agree with each other within $\approx 1\sigma$, for any PDF set. 
The compatibility of the results obtained is a sign of their robustness.
We observe that data related to $t\bar{t}$ decays in the dileptonic channel (Refs.~\cite{top18004,top14013}) point towards central \mtpole values smaller than data related to decays in the semileptonic channel (Refs.~\cite{a190807305,top20001}).
The values extracted from all ATLAS and all CMS differential measurements are compatible within $2.5\sigma$, and the same level of compatibility is observed for the results extracted from the measurements either in the dileptonic or in the semileptonic \ttbar decay channels.
In both cases, the difference arises almost entirely from the CMS measurements of Refs.~\cite{top14013,top18004} which point to a central value of \mtpole lower than all other measurements. More detail is available in Ref.~\cite{Garzelli:2023rvx}.

\section{Conclusions}
\label{sec:conclu}

In summary, adding the differential data to the \mtpole fit only including total \ttbar $+ X$ cross sections plays a crucial role in decreasing the uncertainties on \mtpole by a factor of $\sim$ 3. The result of the most comprehensive fit has an uncertainty band ranging from 0.3 to 0.5~GeV, depending on the PDF set.
These uncertainties are a factor 2.5 smaller than those of the most recent average presented in the PDG~\cite{ParticleDataGroup:2022pth}, $\mtpole=172.5\pm 0.7$~GeV. One should also observe that uncertainties related to the data used have similar size to (scale~+~PDF variation) uncertainty at fixed PDF set. We expect that forthcoming experimental data from Run 3 and Run 4 will seriously challenge theoretical capabilities of reducing theory uncertainties to
a similar level as well. 

Our present work can be regarded as a proof-of-principle that 
a simultaneous fit of \mtpole, PDFs and $\alpha_s(M_Z)$
at NNLO accuracy, considering the correlations among them and
using state-of-the art total and multi-differential $t\bar{t}$ production
data, is within reach. We plan to perform such a fit in a next work, 
upgrading the precision and accuracy of the NLO fit results we presented
in Ref.~\cite{Garzelli:2020fmd}. 

\section*{Acknowledgements}
The work of M.V.G. and S.-O.M. has been partially supported by the Bundesministerium f\"ur Bildung und Forschung (BMBF), under contract 05H21GUCCA. The work of O.~Z. has been supported by the {\it Philipp Schwartz Initiative} of the Alexander von Humboldt foundation.

\bibliographystyle{JHEP}
\bibliography{ttbar}

\end{document}